\documentclass[10pt,aps,article,twocolumn]{revtex4}
\usepackage{amsmath}    
\usepackage{amsfonts} 
\usepackage{amssymb}
\usepackage{latexsym}
\usepackage{graphics}
\usepackage{epsfig}
\usepackage{layout}
\usepackage{amssymb}
\usepackage{amsmath}    
\usepackage{amsfonts} 
\usepackage{latexsym}
\usepackage{graphics}
\usepackage{graphicx}
\usepackage{epsfig}
\usepackage{layout}
\usepackage{pgfplots}
\usepackage{pgfplotstable}
\usepackage{tikz}
\usepackage[utf8]{inputenc}
\usepackage[T1]{fontenc}
\usepackage{lmodern} 
\usepackage{listings}
\usepackage{color}
\usetikzlibrary{arrows}
\selectcolormodel{rgb}
\definecolor{Mygreen}{rgb}{0.15,0.75,0.45}
\pgfplotsset{compat=1.5}
\usepackage{pict2e}
\usepackage{graphicx}
\usepackage[compact]{titlesec}
\titlespacing{\section}{0pt}{2ex}{2ex}

\selectcolormodel{cmyk}
\usepackage{lipsum}
\newcommand{\ii }{\'{\i}}
\usepackage[utf8]{inputenc}
\usepackage[T1]{fontenc}
\usepackage{lmodern} 
\usepackage{listings}
\usepackage{pict2e}
\usepackage{graphicx}
\usepackage[compact]{titlesec}
\titlespacing{\section}{0pt}{2ex}{2ex}

\begin{document}
\title{Electrical imprint effects on far infrared (FIR) transmittance spectrum in PZT ferroelectric films}
\author{H. Vivas C.}
\affiliation{Departamento de F\'{i}sica,
Universidad
Nacional de Colombia, Sede Manizales, A.A. 127,
Col.}
 \email{hvivasc@unal.edu.co} 
\date{\today}
\begin{abstract}
Tunable transmittance response in the $0.1-25$ THz range for a lead  Zirconate Titanate Ferroelectric  film under imprint effects and surface anisotropy is calculated by adapting the classical Landau Devonshire theory and the Rouard's method. Induced electrical field is introduced by modulating the $P-E$ polarization profile, while the dielectric permittivity frequency dependence enters into the formalism by taking into the account the soft phonon mode E(TO1) contribution in the framework of the Drude-Lorentz model. It is found that two optical states of light transmittance emerge at zero applied field and normal incidence, and the intensities of transmitted light are closely correlated with the strength of imprint and the path of the electrical polarization.
\newline
\leftline{DOI:http://dx.doi.org/10.1016/j.optcom.2015.02.011\hspace{2cm}PACS numbers: }
\end{abstract}
\maketitle
\section{Introduction}
Electrical imprint is generally considered as an undesirable effect in FeRAM technology mainly because it attempts against the data storage stability \cite{ben}. Nevertheless, electrical imprint treatments on ferroelectric arrays have risen special interest in the last decade since they have demonstrated a crucial role in the design of the shape of non-volatile memories in piezoelectric actuators \cite{AKT}. Physical origin of imprint is still under debate, although the ferroelectric degradation of polarization properties associated to non-switching surface layers with a large residual field in the electrode-ferroelectric frontier has been identified as one responsible mechanism for the shifting in the hysteresis loop \cite{Abe},\cite{Zhou}. Imprint control can be achieved either by exposing the sample during long periods and high temperatures \cite{Kadota}, by manipulating the thickness of pinned domains on the free lateral surfaces \cite{Alexe} or by injecting electronic charges into the electrode-ferroelectric interface via Schottky thermoionic current \cite{kr}. First procedure has been successfully implemented in experimental lead zirconate titanate (PZT) optical shutters with stable performance on its dielectric susceptibility response after a long range of commutation pulses ($\sim 10^4$)\cite{ohashi1},\cite{ohashi2}, outlining an alternate principle on light transmittance memory devices. Advances in the THz limit technology have also found promising proposals for low power operation on hybrid ferroelectric/graphene layer nanoplasmonic waveguides \cite{Gu},\cite{Jin}. On this scenario, we introduce the electrical imprint strength as an essential mechanism for the observed offset in the characteristic hysteresis $P\left(E\right)$ loop, and calculate the effective index of refraction, the optical transmittance and the shape of memory under typical applied fields up to 300 kV/cm for 800 nm PZT systems in the edge of low THz. It is shown that asymmetric states of light transmittance arise by manipulating the strength of a vertical or horizontal imprint at zero field, in agreement with recent experimental reports \cite{ohashi2}.
\section{Transmittance Response Model}
\begin{figure}[h]
\centering
\begin{tikzpicture}[scale=1,line width=3pt]
\node[rectangle,draw, ultra thick, fill=green!20,
       draw,thick,minimum width=3cm,minimum height=0.75cm] {\hspace{1.5 cm}$\mathbf{P}\left(z,\mathbf{E}\right)$};
\node at (0.0,-0.75) {\large{$\ell$}};
\draw[ultra thick, ->,red] (-1.0,0.0) -- (0.0,0);
\draw[thick, ->,black] (-3*pi/2,0.0) -- (-3*pi/2,0.75);
\draw[thick, ->,black] (-3*pi/2,0.0) -- (-2.250,0.0);
\draw[thick, ->,black] (-3*pi/2,0.0) -- (-2.250,0.0);
\draw[ultra thick, ->,purple] (-3.5,0.5) -- (-3,0.5);
\draw[thick, ->,black] (-3*pi/2,0,0) -- (-3*pi/2,0,1.0);
\draw[thick, <->,black] (-1.5,-0.5) -- (1.5,-0.5);
\node at (-5,-0.70){$Y$};
\node at (-4.35,0.75){$X$};
\node at (-2,0.0){$Z$};
\draw[blue,thick] plot[domain = -3*pi/2:-2.5, samples = 100] 
            (\x,0,{1.0*sin(10*(\x) r)});
 \end{tikzpicture}
\caption{Schematics for an incident electromagnetic plane wave interacting with a ferroelectric sample. The average optical path depends on the external electrical field through the effective index of refraction $\check{n}_{e}$: $OP=\ell\check{n}_{e}\left(\omega,E\right)$.}\label{sqm}
\end{figure}
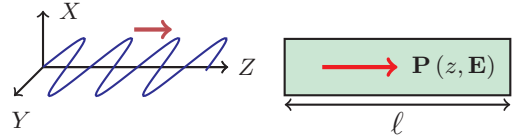

 Figure (\ref{sqm}) sketches the frame set for incident electromagnetic waves in the THz range interacting with a ferroelectric slab with variable index of refraction. Transmitted waves spectra are modified by imprinted and externally electrical field $\mathbf{E}$.  
 The classical Landau-Ginzburg-Devonshire (LGD) model \cite{modelLGD},\cite{FD},\cite{Chandra} provides the description for the polarization field distribution $P\left(z,\mathbf{E}\right)$ in ferroelectric (FE) phase under an applied electrical field $\mathbf{E}$. The single component for the polarization field  $P\left(z,\mathbf{E}\right)\equiv P$ is obtained by solving the third order non-linear differential equation:
\begin{equation}\label{profile}
-\alpha\xi_{b}^{2}\frac{\partial^{2}P}{\partial z^{2}}+\alpha P+\beta P^3 =\mathbf{E\cdot \hat{n}}, 
\end{equation} 
where $\alpha$ and $\beta$ are the typical parameters taken from the renormalized Gibbs free energy functional in the $c$-phase configuration \cite{Khare}, $\xi_b$ corresponds to the correlation length in FE state and $E=\mathbf{E\cdot \hat{n}}$ defines the relative (and uniform) field intensity along $z$-direction. The polarization field distribution profile nearby the surface of a ferroelectric film of thickness $\ell$ changes with its perpendicular distance as 
\begin{equation}\label{profileBC}
\left(\frac{\partial P}{\partial z}\right)_{z=0,\ell}=\pm\lambda^{-1}P\left(z=0,\ell\right), 
\end{equation}
according with the Kretschmer's theory \cite{K}. $\lambda$ encodes the asymmetric depolarization field effects due to a large variety of phenomena, among others, the relative orientation of $\mathbf{P}$ respect to the normal of the surfaces, the proximity vacuum-interface boundary depletion field or the mismatch strain for samples in contact with a substrate \cite{Wang}. In the frame of a linearized approach,  the complete polarization profile is constructed by writing $P\approx\langle P\left(E\right)\rangle+\delta p\left(z,E\right)$, where $\langle P\left(E\right)\rangle$ is taken as the average polarization of the film calculated from (\ref{profile}), and $\delta p\left(z,E\right)\sim\delta p$ corresponds to the spatial fluctuations around $\langle P\left(E\right)\rangle$. Therefore, $\delta p$ must satisfy:
\begin{equation}\label{profileL}
-\alpha\xi_{b}^{2}\frac{\partial^{2}\delta p}{\partial z^{2}}+\bar{\alpha}\delta p= E,
\end{equation} 
with $\bar{\alpha}=\alpha+3\beta\langle P\left(E\right)\rangle^{2}$.
The solution for $P\left(z,E\right)$ with boundary conditions (\ref{profileBC}) is given explicitly by:
\begin{equation}\label{ppee}
P\left(z,E\right)=H\left(E\right)\left(1+F\left(E\right)\cosh{\left(\frac{2z-\ell}{2\bar{\xi}_{b}}\right)}\right),
\end{equation}
where $H\left(E\right)=\langle P\left(E\right)\rangle+\bar{\alpha}^{-1}E$, and the structure factor $F\left(E\right)$ takes into account the characteristic lengths in the system, namely, the thickness $\ell$, the surface depletion $\lambda$ and the renormalized coherence length in the FE state $\bar{\xi}_{b}\left(E\right)\equiv\bar{\xi}_{b}=\xi_{b}\sqrt{\alpha/\bar{\alpha}}$, which depends on the transition temperature $T_{C}$ and the external field intensity. The dependence for the average polarization $\langle P\left(E\right)\rangle$ must be obtained self-consistently \cite{Lu}. Factor $F\left(E\right)$ is calculated as $F^{-1}\left(E\right)=-\left[\cosh{\left(\ell/2\bar{\xi}_{b}\left(E\right)\right)}+\left(\lambda/\bar{\xi}_{b}\left(E\right)\right)\sinh{\left(\ell/2\bar{\xi}_{b}\left(E\right)\right)}\right]$. By inserting the adjustment correlations in terms of the coercive field $E_c$ and the remnant polarization at zero field $P_{r}$ as $\mid\alpha\mid=3\sqrt{3}E_{c}/2P_{r}$ and $P_{r}=\sqrt{\mid\alpha\mid/\beta}$, the expression for the dielectric susceptibility $\chi\left(E\right)$ as an intrinsic function of applied field $E$ in the framework of the Landau-Khalatnikov theory is derived \cite{coercive},\cite{LK}: 
\begin{equation}\label{chimodel}
\chi\left(E\right)=\frac{2P_{r}^{3}G\left(E\right)}{3\sqrt{3}\varepsilon_{0}E_{c}\left(P_{r}^{2}+3\langle P\left(E\right)\rangle^{2}\right)},
\end{equation}
with $G\left(E\right)=1+2\left(\bar{\xi}_{b}\left(E\right)/\ell\right)F\left(E\right)\sinh{\left(\ell/2\bar{\xi}_{b}\left(E\right)\right)}$.  Hence, the inhomogeneous index of refraction $n_{\omega}\left(z,E\right)$ is written as \cite{eliseev},\cite{Vivas}: 
\begin{equation}
n_{\omega}\left(z,E\right)=n\left(\omega\right)\left[1+Q\left(z,E\right)\right]^{1/2},
\end{equation}
with $Q\left(z,E\right)=\chi\left(E\right)\left(1+F\left(E\right)\cosh{\left[\left(2z-\ell\right)/2\bar{\xi}_{b}\left(E\right)\right]}\right)$. The correlation with the phase difference associated to the optical path traveled by a coherent electromagnetic wave for those spatially inhomogeneous systems is calculated through the definition \cite{phased}:
\begin{equation}\label{fase}
\delta=\frac{\omega}{c}\int_{0}^{\ell}n_{\omega}\left(z,E\right)dz=\frac{\omega}{c}\ell\check{n}_{e}\left(\omega,E\right),
\end{equation}
where $\check{n}_{e}\left(\omega,E\right)\equiv\check{n}_{e}$ is taken as the \emph{effective} value for the index of the refraction in the length $\ell$:
\begin{equation}\label{naverage}
\check{n}_{e}=n\left(\omega\right)\kappa_{1}^{-1}\left[1+Q\left(\ell/2,E\right)\right]^{1/2}\mathcal{E}\left(\kappa_{1}\mid\kappa_{2}\right),
\end{equation} 
$\mathcal{E}\left(\kappa_{1}\mid\kappa_{2}\right)$ represents the incomplete elliptic integral of the second kind \cite{elli}  with $\kappa_{1}\equiv\kappa_{1}\left(E\right)=i\ell/4\bar{\xi}_{b}\left(E\right)$ and $\kappa_{2}\equiv\kappa_{2}\left(E\right)=2F\left(E\right)\chi\left(E\right)/\left(1+Q\left(\ell/2,E\right)\right)$. Far infrared dielectric response of Lead (Zirconate) Titanate (PT-PZT) have been intensively investigated by using the classical Drude-Lorentz type model in the THz range \cite{Fedo},\cite{Buix},\cite{Nakada}:
\begin{equation}\label{DL}
\check{\varepsilon}\left(\omega\right)\sim\varepsilon_{\small{\mbox{THz}}}+\frac{\Delta\varepsilon_{\small{\mbox{THz}}}\omega_{M}^{2}}{\omega_{M}^{2}-\omega^{2}+i\gamma\omega};
\end{equation}
where $\omega_{M}\sim 1.6$ THz and $\Delta\varepsilon_{\small{\mbox{THz}}}=150$ correspond to the adjusted frequency of the phononic E(TO1) soft central mode selected for (undoped) PZT films \cite{Li},\cite{Wangg}. Its relationship with the factor $n\left(\omega\right)$ in Equation (\ref{naverage}) is given by $n\left(\omega\right)=\sqrt{\check{\varepsilon}\left(\omega\right)}$. Two Debye relaxation mechanisms are present in the permittivity response model, however, they lie into the GHz range and do not have significant effects on the transmittance spectrum in the interval of interest. 
The transmittance response  $T\left(\omega,E\right)\equiv T_{\omega}=\mbox{abs}\left[\check{\tau}\right]^{2}$ is calculated by using the Rouard's method for normal incidence as a function of the phase difference $\delta$, the film index $\check{n}_{e}$ and its environment $n_0$ \cite{Lecaruyer},\cite{Heavens}:
\begin{equation}\label{tauT}
\check{\tau}=\frac{\check{t}\check{t}^{\prime}e^{-i\delta}}{\left(1-\check{r}^{2}e^{-2i\delta}\right)},\hspace{1cm} \check{t}=\frac{2}{\left(1+\check{n}_{e}/n_{0}\right)},
\end{equation}
 $\check{t}^{\prime}=\left(\check{n}_{e}/n_{0}\right)\check{t}$ and $\check{r}=\left(1-\check{n}_{e}/n_{0}\right)\left(1+\check{n}_{e}/n_{0}\right)^{-1}$.  Equations (\ref{fase})-(\ref{tauT}) constitute the core results in this paper: the transmittance spectrum due to the propagation of electromagnetic (THz) radiation in a ferroelectric film under vertical and horizontal imprint effects. Numerical results are discussed in the next section.
\section{Results}
\begin{figure} 
\centering
\pgfplotsset{every axis/.append style={
extra description/.code={
\node at (0.85,0.65) {$x=+0.5$};
\node at (0.90,0.925) {$y=+0.5$};
\node at (0.95,0.25) {$E\times$10$^{2}$ (kV/cm)};
\node at (0.60,0.8) {$\textbf{\emph{I}}$};
\node at (0.80,0.85) {$\textbf{\emph{II}}$};
\node at (0.75,0.550) {$\textbf{\emph{III}}$};
\node at (0.15,0.08) {$\textbf{\emph{IV}}$};
\node at (0.425,0.48) {$\textbf{\emph{V}}$};
}}}
\begin{tikzpicture}[trim axis left, trim axis right]
\begin{axis}[title=$P\left(E\right)\times 10^{2} \left(\mu \mbox{C/cm}^{2}\right)$,
axis lines=middle,
axis line style={->}]
\addplot [domain=-5.0:5.0,black,thick,smooth] {0.3738*tanh(0.56404*(1.2636*x+1))};
\addplot [domain=-5.0:5.0,black,thick,smooth] {0.3738*tanh(0.56404*(1.2636*x-1))};
\addplot [domain=-5.0:5.0,blue,thick,smooth] {0.3738*tanh(0.56404*(1.2636*x+1-0.5))};
\addplot [domain=-5.0:5.0,blue,thick,smooth] {0.3738*tanh(0.56404*(1.2636*x-1-0.5))};
\addplot [domain=-5.0:5.0,red,thick,smooth] {0.5*0.3738+0.3738*tanh(0.56404*(1.2636*x+1))};
\addplot [domain=-5.0:5.0,red,thick,smooth] {0.5*0.3738+0.3738*tanh(0.56404*(1.2636*x-1))};
\draw[->,thick,black] (axis cs:0.05,0.252) -- (axis cs:0.5,0.32);
\draw[->,thick,black] (axis cs:3.5,0.4) -- (axis cs:4.5,0.4);
\draw[->,thick,black] (axis cs:-3.5,-0.33) -- (axis cs:-4.5,-0.33);
\draw[->,thick,black] (axis cs:0.95,0.1) -- (axis cs:0.6,0.01);
\end{axis}
\end{tikzpicture}
\caption{(Color)$P-E$ model response for PT ferroelectric sample with $E_{c}=79.14$ kV/cm, $P_{r}=19.10 \mu$C/cm$^2$, $P_{s}=37.38 \mu$C/cm$^2$. Electrical imprint effects are shown for $x=+0.5$ (blue line) and $y=+0.5$ (red line). Full set of parameters are taken as $\ell=800$ nm, $\lambda/\ell=0.12$, $\xi_{b}/\ell=0.1.$}\label{hys}
\end{figure}
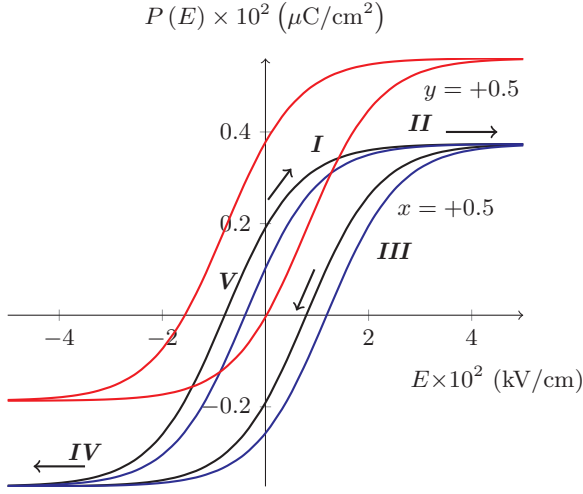
\begin{figure}
\includegraphics[width=1.0\columnwidth,keepaspectratio]{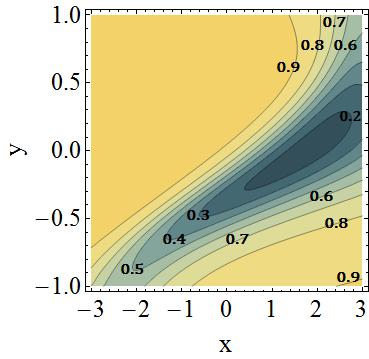}
\caption{Contour maps for the transmittance response under vertical ($y$) and horizontal ($x$) imprint effects in \emph{inverse} polarization regime at 0.1 THz and null applied field.}\label{coni}
\end{figure}

Electrical polarization induced by imprint treatment has been qualitatively set into the formalism in terms of the saturation $P_{s}$ and coercive fields under the adjustment $\langle P_{\pm}\left(E\right)\rangle\equiv yP_{s}+P_{s}\tanh{\left[A\left(E\pm E_{c}-xE_{c}\right)\right]}$ in Eq. (\ref{chimodel}), with $A=\tanh^{-1}{\left(P_{r}/P_{s}\right)}/E_{c}$ \cite{Tag},\cite{Grossmann},\cite{NI}. Parameter $y$ might encode the relative vertical imprint strength, and its negative value indicates that imprinting treatment has been performed by inducing an external polarization field in opposite direction relative to $P_r$.  Factor $x$ represents the horizontal shift taken as a fraction of the coercive electrical field $E_c$. Figure (\ref{hys}) compares the hysteresis profile for a symmetrical loop without imprint treatment (black line), its vertical shifting under imprint effects for $y =+0.5, x=0$ (red line) and the characteristic bias for $y=0, x=+0.5$ (blue line). Five regimes are readily identified depending on the sign in $\langle P_{\pm}\left(E\right)\rangle$: ($I$) direct polarization for positive increasing field, ($II$) saturation at $+P_{r}$, ($III$) depolarization, ($IV$) reverse saturation at $-P_{r}$ and ($V$) direct polarization for negative increasing field.
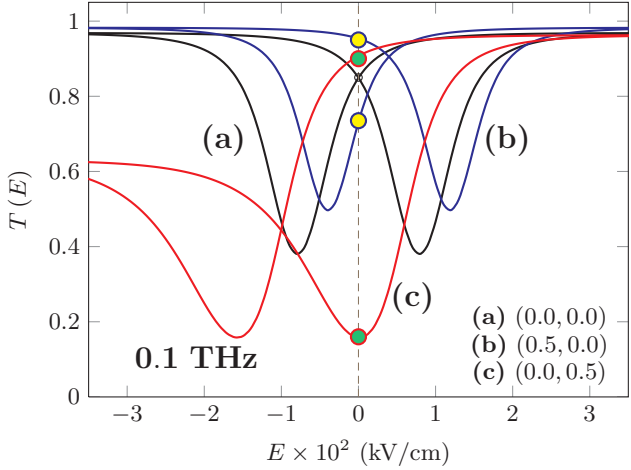
\begin{figure}
\centering
\pgfplotsset{every axis/.append style={
extra description/.code={
\node at (0.2,0.10) {\large{$\mathbf{0.1}$ \bf{THz}}};
\node at (0.25,0.65) {\large{\textbf{(a)}}};
\node at (0.775,0.65) {\large{\textbf{(b)}}};
\node at (0.60,0.25) {\large{\textbf{(c)}}};
\node at (0.835,0.20) {\textbf{(a)} $\left(0.0,0.0\right)$};
\node at (0.835,0.13) {\textbf{(b)} $\left(0.5,0.0\right)$};
\node at (0.835,0.06) {\textbf{(c)} $\left(0.0,0.5\right)$};
}}}
\begin{tikzpicture}[trim axis left, trim axis right]
\begin{axis}[x=1.0250cm,y=5cm,domain=-3.5:3.5,xmax=3.5, xmin=-3.5,ymax=1.05, ymin=0.0,
xlabel={$E\times 10^{2}$ (kV/cm)},
ylabel={$T\left(E\right)$}
]
\addplot[black,thick,smooth] table {thyp5.txt};
\addplot[black, thick,smooth] table {thyp6.txt};
\addplot[blue,thick,smooth] table  {thyp7.txt};
\addplot[blue,thick,smooth] table {thyp8.txt};
\addplot[red,thick,smooth] table  {thyp9.txt};
\addplot [red,thick,smooth] table {thyp10.txt};
\addplot+[ycomb] plot coordinates{(0,1.1)};
\draw[thin, black](axis cs:0.0,0.85) circle (0.05cm);
\draw[thick, blue,fill=yellow](axis cs:0.0,0.95) circle (0.1cm);
\draw[thick, blue,fill=yellow](axis cs:0.0,0.735) circle (0.1cm);
\draw[thick, red,fill=Mygreen](axis cs:0.0,0.9) circle (0.1cm);
\draw[thick, red,fill=Mygreen](axis cs:0.0,0.16) circle (0.1cm);
\end{axis}
\end{tikzpicture}
\caption{$\left(x,y\right)$ imprint effects on $T\left(E\right)$ spectra at 0.1 THz}\label{r2}
\end{figure}
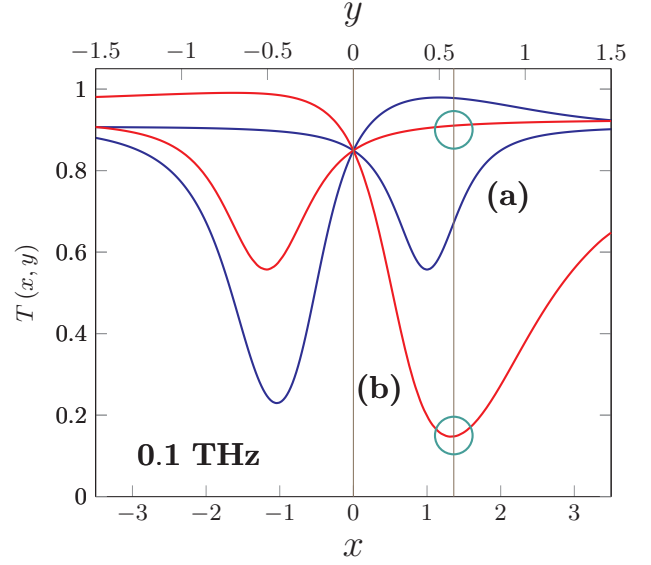
\begin{figure}
\centering
\pgfplotsset{every axis/.append style={
extra description/.code={
\node at (0.2,0.10) {\large{$\mathbf{0.1}$ \bf{THz}}};
\node at (0.80,0.70) {\large{\textbf{(a)}}};
\node at (0.55,0.25) {\large{\textbf{(b)}}};
}}}
\begin{tikzpicture}[trim axis left, trim axis right]
\begin{axis}[
scale=1.0,
domain=-3.5:3.5,xmax=3.5, xmin=-3.5,ymax=1.05, ymin=0.0,
axis x line*=bottom,
xlabel={\Large{$x$}},
ylabel={$T\left(x,y\right)$}
]
\addplot[blue,thick,smooth] table {thyp1.txt};
\addplot[blue,thick,smooth] table {thyp2.txt};
\end{axis}
\begin{axis}[
scale=1.0,
domain=-1.5:1.5,xmax=1.5, xmin=-1.5,ymax=1.05, ymin=0.0,
axis x line*=top,
xlabel={\Large{$y$}},
]
\addplot[red,thick,smooth] table {thyp3.txt};
\addplot[red,thick,smooth] table {thyp4.txt};
\addplot+[ycomb] plot coordinates{(0,1.1) (0.5842,1.1)};
\draw[thick, Mygreen](axis cs:0.5842,0.9) circle (0.25cm);
\draw[thick, Mygreen](axis cs:0.5842,0.15) circle (0.25cm);\end{axis}
\end{tikzpicture}
\caption{ Transmittance spectra dependence at 0.1 THz with $E=0$ as a function of the horizontal $T\left(x,0\right)$ (curve a) and vertical $T\left(0,y\right)$ (curve b) imprint strengths.}\label{mem01}
\end{figure}
Numerical simulations for the hysteresis loop and dielectric susceptibility were performed for PZT films with approximated parameters reported in reference \cite{ZZ}. Figure (\ref{r2}) (a) depicts the relation $T\left(E\right)$ at $0.1$ THz and $x=y=0$. When non-imprinting procedure is taken into account, the hysteresis loop remains symmetrical around $P=0$ axis, and the bi-state optical transmittance mode vanishes (single black dot in line (a) at $E=0$). The spectrum exhibits a symmetrical \emph{butterfly shape} with minimal transmittance peaks at the coercive fields $\pm E_{c}$. In the case $x=+0.5, y=0$ (curve (b)), the spectrum is horizontally biased and two states of light transmittance emerge at zero field depending on the path of the polarizability ($I$) or ($III$). The minima of $T\left(E\right)$ remain symmetrical around its crossing point (identified as the electrical field value for which the transmittance function gets the same value for direct and inverse polarization), with higher values compared with the case (a). Line (c) is calculated for $x=0$ and $y=+0.5$. The crossing point is shifted for negative values of the externally applied field, the butterfly-shape symmetry breaches and the spectrum shows its maximum difference for the transmittance levels at $E=0$, indicating that a stronger optical response arises when the \emph{vertical} imprint treatment predominates. 
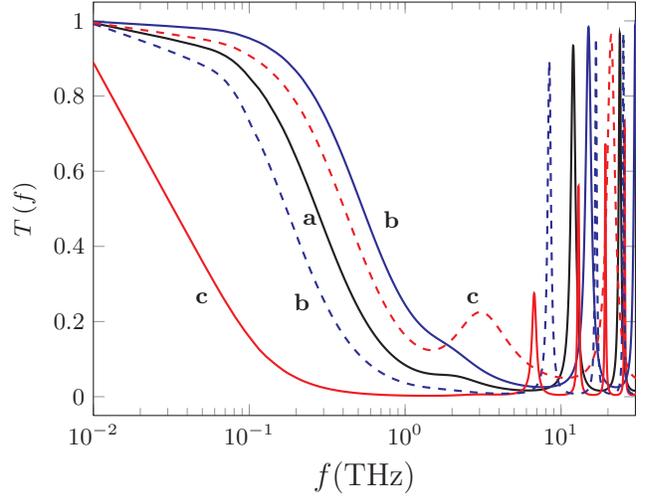
\begin{figure}
\centering
\pgfplotsset{every axis/.append style={
extra description/.code={

\node at (0.70,0.285) {\small{\textbf{c}}};
\node at (0.20,0.285) {\small{\textbf{c}}};
\node at (0.385,0.275) {\small{\textbf{b}}};
\node at (0.550,0.475) {\small{\textbf{b}}};
\node at (0.4,0.475) {\small{\textbf{a}}};
}}}
\begin{tikzpicture}[trim axis left, trim axis right]
\begin{semilogxaxis}[x=0.9cm,y=5cm,domain=0:30,xmax=30, xmin=0.01,ymax=1.05, ymin=-0.05,
xlabel={\large{$f$(THz)}},
ylabel={$T\left(f\right)$}
]
\addplot[black, thick,smooth] table[x index=0,y index=1,col sep=tab]{FIR222.txt};
\addplot[blue, thick,smooth] table[x index=0,y index=2,col sep=tab]{FIR222.txt};
\addplot[red, thick,smooth] table[x index=0,y index=3,col sep=tab]{FIR222.txt};
\addplot[blue, thick,smooth,dashed] table[x index=0,y index=2,col sep=tab]{FIR22.txt};
\addplot[red, thick,smooth,dashed] table[x index=0,y index=3,col sep=tab]{FIR22.txt};
\end{semilogxaxis}
\end{tikzpicture}
\caption{Transmittance spectrum as a function of the incident wave frequency for several imprint strength at zero field. (a) $x=y=0$. (b) $x=0.5, y=0$. (c) $x=0, y=0.5$. Solid (dashed) Lines correspond to the points evaluated at $E=0$ in the path \emph{I} (\emph{III}) on the hysteresis curve Figure (\ref{hys}).}\label{mem02}
\end{figure}
\begin{figure}
\includegraphics[width=1.0\columnwidth,keepaspectratio]{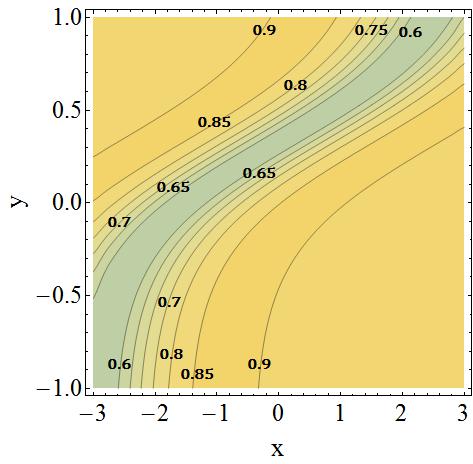}
\caption{Contour maps for the transmittance response under vertical ($y$) and horizontal ($x$) imprint effects in \emph{direct} polarization regime at 0.1 THz and null applied field.}\label{cond}
\end{figure}
Figure (\ref{mem01}) shows the evolution of the transmittance spectrum at fixed frequency (0.1 THz) for independent values of $x$ and $y$ in the range between $\lbrace -3.5;3.5\rbrace$ and $\lbrace-1.5;1.5\rbrace$ respectively, with the largest difference around 75$\%$ (labeled in open circles) at $y\sim 0.58$. Figure (\ref{mem02}) illustrates the transmittance for different imprint strengths from $\sim 0.01$ up to 25 THz in the far infrared edge (FIR) for direct and inverse polarization. Absorption associated to the E(TO1) phonon mode becomes apparent under vertical imprint for $\sim 1-5$ THz (line c) and is essentially absent for horizontal bias. In the range of higher frequencies (>10 THz), the film exhibits peaks of transparency whose positions depend on the values of $x$ and $y$. The transmittance overlaps their lines regardless its polarization bias for $x=y=0$ (curve a), but it takes two values depending on the cycle history when imprint effects are included into the calculations.    Figures (\ref{cond}) and (\ref{coni}) show the contour lines in the transmittance response for simultaneous $\left(x,y\right)$ imprint strengths in two different paths of polarization. On the process $V\rightarrow I\rightarrow II$ the response changes monotonously from 0.65 to 0.9 in contrast with remarkable variations after calculating it upon the $III\rightarrow IV$ trajectory. Optical transmittance differences at zero field $\Delta T=\mid T_{V\rightarrow I\rightarrow II}-T_{III\rightarrow IV}\mid$ for direct and inverse polarization states as a function of the positive imprint strengths $\left(x,y\right)$ and various frequencies in the THz regime are shown in Figure (\ref{dTX}). Non \emph{shape memory effect} is available without imprinting treatment ($y=x=0$). $\Delta T$ response is sensitive to the external radiation frequency since it tends to increase as the frequency approaches to the edge of the far infrared regime ($\sim 0.1$ THz). The E(TO1) phonon mode contribution becomes significant in the range between 1 to 3 THz, as referred in the $0<y<0.3$-crossover. 
\begin{figure}
\centering
\pgfplotsset{every axis/.append style={
extra description/.code={
\node at (0.4,0.60) {\textbf{0.1 THz}};
\node at (0.65,0.75) {\textbf{0.5 THz}};
\node at (0.4,0.10) {\textbf{3.0 THz}};
\node at (0.7,0.30) {\textbf{1.0 THz}};
\node at (0.5,-0.15) {\large{$x$}};
\node at (0.0,1.1) {\large{$\Delta T\left(x\right)$}};
}}}
\begin{tikzpicture}[trim axis left, trim axis right]
\begin{axis}[scale=1.0,x=2.5cm,y=6.5cm,domain=0.0:3.0,xmax=3.0, xmin=0.0,ymax=0.8, ymin=0.0,
axis lines=middle,
axis line style={->},legend pos= north east]
\addplot[red, thick,smooth] table[x index=0,y index=1,col sep=tab]{deltaXXX.txt};
\addplot[blue, thick,smooth] table[x index=0,y index=2,col sep=tab]{deltaXXX.txt};
\addplot[Mygreen, thick,smooth] table[x index=0,y index=3,col sep=tab]{deltaXXX.txt};
\addplot[black, thick,smooth] table[x index=0,y index=4,col sep=tab]{deltaXXX.txt};
\end{axis}
\end{tikzpicture}
\end{figure}
\begin{figure}
\centering
\pgfplotsset{every axis/.append style={
extra description/.code={
\node at (0.60,0.60) {\textbf{0.5 THz}};
\node at (0.65,0.95) {\textbf{0.1 THz}};
\node at (0.60,0.175) {\textbf{1.0 THz}};
\node at (0.60,0.425) {\textbf{3.0 THz}};
\node at (0.5,-0.15) {\large{$y$}};
\node at (0.0,1.1) {\large{$\Delta T\left(y\right)$}};
}}}
\begin{tikzpicture}[trim axis left, trim axis right]
\begin{axis}[scale=1.0,x=5.0cm,y=6.5cm,domain=0.0:1.5,xmax=1.5, xmin=0.0,ymax=0.8, ymin=0.0,
axis lines=middle,
axis line style={->},legend pos= north east]
\addplot[red,thick,smooth] table {thypv10.txt};
\addplot[blue,thick,smooth] table {thypv11.txt};
\addplot[Mygreen,thick,smooth] table {thypv12.txt};
\addplot[black,thick,smooth] table {thypv13.txt};
\end{axis}
\end{tikzpicture}
\caption{Shape memory effects as a function of $\left(x,y\right)$-imprint and different frequencies in the THz range, with zero applied field.}\label{dTX}
\end{figure}
\begin{figure}
\centering
\pgfplotsset{every axis/.append style={
extra description/.code={
\node at (0.225,0.55) {\small{(\textbf{c$^{\prime}$})}};
\node at (0.15,0.285) {\small{(\textbf{b$^{\prime}$})}};
\node at (0.1,0.075) {\small{(\textbf{a$^{\prime}$})}};
\node at (0.225,0.18) {\small{(\textbf{d$^{\prime}$})}};
}}}
\begin{tikzpicture}[trim axis left, trim axis right]
\begin{axis}[x=1.35cm,y=5.25cm,domain=0:5.2,xmax=5.2, xmin=-0.1,ymax=0.85, ymin=-0.05,
xlabel={\large{$f$(THz)}},
ylabel={$\Delta T\left(x,0\right)$}
]
\addplot[black, thick,smooth] table[x index=0,y index=1,col sep=tab]{deltaXX.txt};
\addplot[blue, thick,smooth] table[x index=0,y index=2,col sep=tab]{deltaXX.txt};
\addplot[red, thick,smooth] table[x index=0,y index=3,col sep=tab]{deltaXX.txt};
\addplot[Mygreen, thick,smooth] table[x index=0,y index=4,col sep=tab]{deltaXX.txt};
\end{axis}
\end{tikzpicture}
\caption{$\Delta T\left(x,0\right)$ for $E=0$ and (a$^{\prime}$) $x=0.1$, (b$^{\prime}$) $x=0.5$, (c$^{\prime}$) $x=1.0$, (d$^{\prime}$) $x=3.0$. }\label{imp1}
\end{figure}
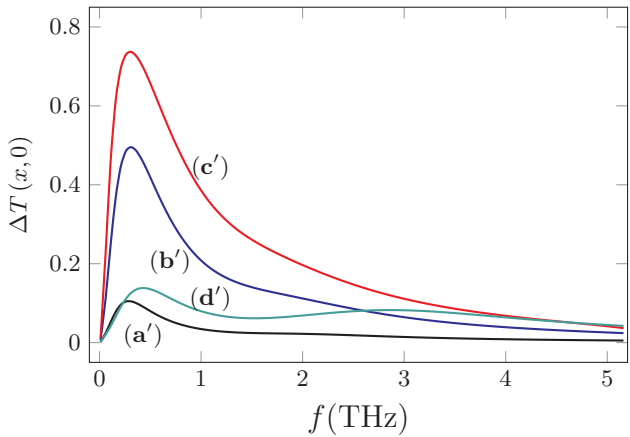
\begin{figure}
\centering
\pgfplotsset{every axis/.append style={
extra description/.code={
\node at (0.80,0.55) {\small{(\textbf{c})}};
\node at (0.8,0.285) {\small{(\textbf{b})}};
\node at (0.8,0.125) {\small{(\textbf{a})}};
\node at (0.9,0.4) {\small{(\textbf{d})}};
}}}
\begin{tikzpicture}[trim axis left, trim axis right]
\begin{axis}[x=1.35cm,y=5.25cm,domain=-0.1:5.2,xmax=5.2, xmin=-0.1,ymax=0.85, ymin=-0.05,
xlabel={\large{$f$(THz)}},
ylabel={$\Delta T\left(0,y\right)$}
]
\addplot[black, thick,smooth] table[x index=0,y index=1,col sep=tab]{deltaTT.txt};
\addplot[blue, thick,smooth] table[x index=0,y index=2,col sep=tab]{deltaTT.txt};
\addplot[red, thick,smooth] table[x index=0,y index=3,col sep=tab]{deltaTT.txt};
\addplot[Mygreen, thick,smooth] table[x index=0,y index=4,col sep=tab]{deltaTT.txt};
\end{axis}
\end{tikzpicture}
\caption{$\Delta T\left(0,y\right)$ for $E=0$ and (a) $y=0.1$, (b) $y=0.5$, (c) $y=1.0$, (d) $y=1.5$.}\label{imp2}
\end{figure}
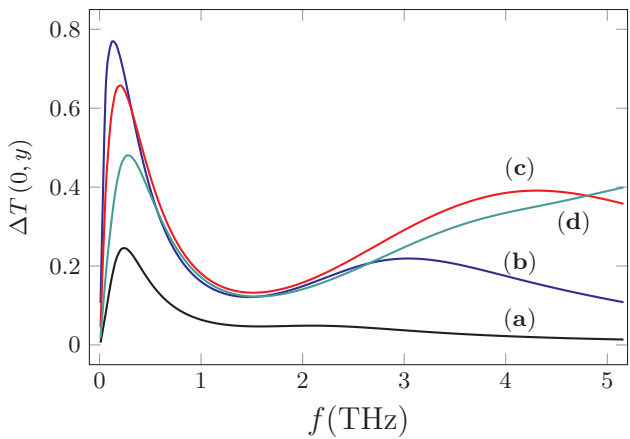
Figures (\ref{imp1}) and (\ref{imp2}) describe the shape of the memory as a function of the external frequency for horizontal and vertical imprinting shift, respectively. The maximum of $\Delta T\left(x,0\right)$ lies at $x=1.0$ (line c$^{\prime}$ in Fig. (\ref{imp1})) and $y\sim 0.5$ (line b in Fig. (\ref{imp2})) for frequencies below 0.1 THz, as already demonstrated in Figure (\ref{mem01}). The effective index of refraction is modified by imprint procedures and phononic excitations for higher frequencies, affecting the transmittance functionality by diminishing its response as long as the frequency reaches the 5 THz limit.
\section{Conclusions}
The explicit relationships for the effective index of refraction $\check{n}_{e}$ and the transmittance response $T_{\omega}$ in ferroelectric (PZT) films with surface anisotropy and induced electrical imprint are calculated by recasting the LGD model and Rouard's technique. It is shown that the transmittance spectrum is highly sensible under imprint strength in the edge of Terahertz range (0.1-1 THz) and depolarizing regime for 800 nm samples. Our approach is solely focused on the vertical and horizontal $\langle P\left(E\right)\rangle$ hysteresis loop displacement, although the model might be directly extended for asymmetrical slanted loops, resembling recent experiments reported for PZT films under Ba$^{+2}$ (Sr$^{+2}$) modifications of dopant concentration in Pb sites \cite{Ng}, broadening a wide set of possibilities in electrochemical control on remnant polarization, coercive field and piezoelectric response in these ceramic materials.  $\langle P\left(E\right)\rangle$ in equation (\ref{ppee}) might be solved exactly, the line becomes slightly different compared with the proposed hyperbolic profile, but this procedure does not change in significant way the main behavior on the transmittance response. Detailed studies on the role of Zr/Ti compositional variation in PZT films have also demonstrate a close correlation between critical temperature $T_{C}$, the short-long structural order crossover passing through rhombohedral-morphotropic phase boundary (MPB)-tetragonal phases, and piezoelectric activity in Lead Titanate system \cite{Jaffe},\cite{Noheda},\cite{Glazer},\cite{DJF}, recalling the pertinence of LGD model for transitional states, specifically in its characteristic length $\bar{\xi}_{b}\rightarrow\bar{\xi}_{b}\left(T_{C}\right)$. Transmittance measurements in ferroelectric thin film structures allow to perform indirect adjustments to Drude-Lorentz model (\ref{DL}) on its soft mode $\omega_{M}\rightarrow \omega_{M}\left(E\right)$ and damping $\gamma\rightarrow\gamma\left(E\right)$ parameters as the dielectric permittivity changes up to 10\% for $E\sim 100$ kV/cm \cite{Kuzel1} and 65\%  and 67 kV/cm for strained samples \cite{Kuzel2}. Long-lasting imprint exposure in ferroelectric films, their intrinsic loss of polarization and retention effects might acquire relevance in highly confined systems with $\ell <50$ nm, and they constitute interesting issues that shall be considered on further investigations. 
\section{Acknowledgements}
The author acknowledges availability and technical support at CER Computer Lab. 

\end{document}